\documentclass[fleqn,twoside]{article}%
\usepackage{amsmath}
\usepackage{amsfonts}
\usepackage{amssymb}
\usepackage{graphicx}
\usepackage{caption}
\usepackage{cite}

\def\be{\begin{equation}}
\def\ee{\end{equation}}
\def\bi{\bibitem}
\begin{document}
\title{Phase space structure of symmetric teleparallel theory of gravity.}
\author{Dalia Saha$^1$, Abhik kumar sanyal$^2$ }
\maketitle
\noindent
\begin{center}
\noindent
$^{1,2}$ Dept. of Physics, Jangipur College, Murshidabad, West Bengal, India - 742213,\\
$^{1}$ Dept. of Physics, University of Kalyani, Nadia, West Bengal, India - 741235.\\
$^2$ Calcutta Institute of Theoretical Physics, Bignan Kutir, 4/1, Mohanbagan Lane, Kolkata, India - 740004.\\
\end{center}
\footnotetext[1]{
\noindent Electronic address:\\
$^1$daliasahamandal1983@gmail.com\\
$^2$sanyal\_ak@yahoo.com\\}
\begin{abstract}
The `Generalized Symmetric Teleparallel Gravity' (GSTG) does not admit diffeomorphic invariance, since the auxiliary field as well as the shift vector act as non-propagating dynamical variables carrying 1/2 degrees of freedom each. We show that in a minisuperspace model, which is devoid of the shift vector, the problem is alleviated for locally Lorentz invariant GSTG theory, and diffeomorphic invariance is established at least for one connection. However, the eerie structure of the Hamiltonian constructed even in the background of spatially flat isotropic and homogeneous Robertson-Walker space-time, can not be maneuvered. In contrast, the other two spatially flat connections containing an arbitrary time dependent function, doesn't admit non-linear extension to `Symmetric Teleparallel Equivalent to General Relativity (STEGR). We therefore construct the phase-space structure with three different spatially flat connections for the `Lorentz invariant' linear-scalar-vector-tensor GSTG action. Diffeomorphic invariance is established and the associated Hamiltonians are found to be well behaved for all the three cases.
\end{abstract}

\section{\bf{Introduction:}}

Teleparallel gravity theory, originally suggested by Einstein to unify gravity and electro-magnetism, has garnered immense interest in recent years. In the theory of distant parallelism or teleparallelism, space-time is characterized by a curvature free linear connection in conjunction with metric tensor field, both defined in terms of a set of four vector fields, called the dynamical tetrad field. The tetrad field was originally introduced to allow distant comparison of the direction of tangent vectors at different points of the manifold, hence it is called distant parallelism. More precisely, since the Riemann tensor vanishes, the parallel transport defined by the covariant derivative $\nabla$ and its associated affine connection $\Gamma_{\mu\nu}^\alpha$ is independent of the path and therefore the terminology `teleparallel' is used \cite{2}. Earlier attempt of Einstein failed since the field equation did not admit Schwarzschild solution. However, the teleparallel gravity as an alternative to the dark-energy issue, resurrected more than a decade back. This theory with torsion is also called the `metric teleparallel gravity'.\\

Einstein's general theory of relativity (GTR) is built out of the torsion-free Levi-Civita connection, which preserves the Riemannian metric. The field equations of GTR is derived from the Einstein-Hilbert (EH) action (in the unit $c = 1$)
\be \label{EH} A_{EH} =  \int {R\over 16\pi G} \sqrt{-g} d^4 x + \mathcal{S}_m,\ee
where, $R$ is the curvature scalar and $\mathcal{S}_m$ is the matter action. A generalized version, in which $R$ is replaced by an arbitrary function of $R$, viz., $f(R)$, called the modified theory of gravity, was proposed to combat the cosmic puzzle (recent accelerated expansion). In the metric teleparallel gravity theory, the Ricci scalar is replaced by the Torsion scalar $\mathrm{T}$ which results in `teleparallel equivalent of general relativity', being abbreviated as TEGR. The reason being, it is essentially GTR apart from a boundary term \cite{A1}. Hence a generalized version had been presented in which an arbitrary function of the torsion scalar $f(\mathrm{T})$ was introduced in the action that could act as an alternative to the dark energy. This generalized $f(\mathrm{T})$ gravity with torsion is also known as the `Generalized Metric Teleparallel theory of Gravity' (GMTG), which has largely been explored from different angles apart from late-time cosmic acceleration \cite{B1,C1} also see \cite{D1} and references therein. In particular, it was shown that conformal gravity with torsion leads to a power law accelerated solution in the matter-dominated era and also the solution is identical to the $\Lambda\mathrm{CDM}$ model \cite{Bam}. Further, as an alternative to the dark-energy issue, GMTG also addresses a more fundamental problem, such as a valid energy-momentum tensor for the gravitational field \cite{2,3,4}. Nevertheless, some serious issues with metric teleparallel gravity soon emerged. Earlier, it was found that, the actions and the field equations are not invariant under local Lorentz transformations (LLT) \cite{LLT}. Although it has been pointed out that LLI might not necessarily be detrimental, especially for teleparallel theories with local Lorentz symmetry \cite{5}, it may be mentioned that LLI implies that the laws of physics remain unaltered for all observers irrespective of their position and momentum. As a result, the choice of time and space axes remains unimportant. The debate associated with extra degrees of freedom in teleparallel theories may also be resolved following a LLI action. Thus, LLI is an issue of consistency of the theory. Indeed, additional degrees of freedom in comparison to GTR have been found to be present which consequently give rise to the strong coupling issues \cite{fT1a,fT1b,fT1c,fT1d,fT1e,fT1f,fT1g,fT1h,fT1i,fT1j,fT1k,fT1l}. The very presence of ghost degrees of freedom (negative norm states) was also reveled soon \cite{fT2}. There is yet another issue. Since Einstein-Cartan theory does'nt propagate in vacuum and the coupling constant is roughly of the order of Planck's scale which only affects Fermions, so torsion is difficult to put to the test. These facts initiated to consider yet another theory known as the `symmetric teleparallel theory of gravity', in which the non-metricity scalar $Q$ (the scalar formed from the rank three tensor $Q_{\lambda\alpha\beta}$, which is the covariant derivative of the metric tensor, and indicates deviation from metricity) replaces the Ricci scalar in the EH action \cite{C}. Nonetheless, this theory again results in GTR apart from a boundary term and is dubbed as `symmetric teleparallel equivalent of general relativity' or STEGR in short. Therefore as before, a generalized version viz., $f(Q)$ is addressed in the action and this theory again plays a key role to replace the dark-energy issue \cite{D}. This generalized version of teleparallel gravity, called the `Generalized Symmetric Teleparallel theory of Gravity' (GMTG) has also been extensively worked out from different perspectives, including the issue of late-time cosmic acceleration \cite{fQa,fQb,fQc,fQd,fQe,fQf}. It may be mentioned that the non-metricity is essentially the failure of the metric tensor to remain covariantly conserved. Therefore, both the magnitudes and angles of vectors and tensors alter while parallel transported. Thus, in contrast to the torsion, non-metricity may be experimentally measured by tracking the trajectory of a test particle. Unfortunately, this theory has also been found to suffer from some serious issues such as strong coupling problem and break down of linear perturbation theory around flat maximally symmetric background \cite{fT2,fQ1,fQ2,fQ3,fQ4}. We shall be back to these issues later again. Our current motivation is to explore yet another important issue in connection with the phase-space structure which has been overlooked so far.\\

Let us now brief the purpose of the present work. In the absence of a viable theory of quantum gravity, quantum cosmology was formulated to extract some facets regarding the nature of gravity in the Planck's epoch. This requires the formulation of the phase-space structure as forerunner. Further, `Inflation' in view of teleparallel theories has already been studied quite extensively, of-course in some specific models \cite{TQInf1, TQInf2, TQInf3, TQInf4, TQInf5, TQInf6, TQInf7, TQlnf8} and excellent agreement with the latest observational data \cite{Planck1, Planck2} has been revealed. However, inflation occurred nearly at the Planck's epoch $(t_P = 10^{-43}~s)$, soon after gravity becomes classical, while all other existing fields remain quantized. This era is described by `quantum field theory in curved space-time' (QFT in CST), in which geometry is treated as classical, while the expectation value of all other fields appear in the energy-momentum tensor. Therefore, inflation is essentially a quantum theory of perturbation. Nonetheless, inflation is usually studied in view of classical field equations, while if a quantum theory of gravity (quantum cosmology to be precise) transits to classical universe under an appropriate semiclassical approximation, then only classical field equations suffice to study inflation. To be precise, if the integrand in the exponent of the semiclassical wave-function is imaginary, then the behaviour of the approximate wave function is oscillatory, and falls within the classically allowed region, otherwise it is classically forbidden. In this context, the well-known `Hartle criterion' for the selection of classical trajectories states that ``If the (semiclassical) wave function of the universe is strongly peaked, then it admits correlations among the geometrical and matter degrees of freedom, and therefore the emergence of classical trajectories of the universe is expected, otherwise, such correlations are lost" \cite{Hartle}. Thus, unless a quantum theory with teleparallel gravity is constructed and its semiclassical wave-function is explored, study of inflation in view of classical field equations is contentious. Construction of the phase-space structure is a precursor to quantizing the theory. In this manuscript, we therefore aim at constructing the phase-space structure of the Hamiltonian, in the background of a minisuperspace model of cosmological interest. It is revealed that although the Hamiltonian constructed using Dirac-Bergmann algorithm is invariant under diffeomorphism, it is awesome and its quantum version would be dreadful for LLI action. Therefore, study of inflation in view of classical field equations is contentious. In contrast, the two other connections containing an arbitrary function $\gamma(t)$ are applicable only to STEGR or to its nonminimal extension. Considering LLI linear-scalar-vector-tensor form of GSTG action, DI is established and the phase-space structure appears quite handy for connection-1, while for connection-2 and 3, DI may only be established constraining the unknown parameter $\gamma(t) \propto N(t)$ and $\gamma(t) \propto N^{-1}(t)$, where $N(t)$ is the lapse function. The Hamiltonians so obtained, although cumbersome, are well behaved and manageable.\\

We organize the current manuscript as follows. In the following section, we discuss the geometry associated with symmetric teleparallel theory of gravity and present the associated covariant form of the field equations for GSTG theory $f(Q,\Gamma)$, treating both the non-metricity scalar and the connection as independent variables. Next we consider isotropic and homogeneous Robertson-Walker (RW) metric in spherical-polar coordinate system, for which connections are non-vanishing and four different non-metricity scalar may be constructed, three in the spatially flat space and one in the curved space. In the present article, we consider spatially flat space only, and in the following three subsections (2.1, 2.2, 2.3) we prove that apart from the first, the two other connections lead to GTR, or to linear-scalar-vector-tensor GSTG action. We therefore present the covariant form of the scalar-vector-tensor theory of GSTG. In section 3, we first construct the phase-space structure of the Hamiltonian for the covariant scalar-tensor action of GSTG theory. Although, DI is established, the phase-space structure of the Hamiltonian is not manageable. Next, we consider covariant scalar-vector-tensor form of the action and formulate phase-space structure with all the three connections for spatially flat RW metric. For the first connection DI is established and the Hamiltonian is found to be simple and well-behaved. For the second and third connections on the contrary, additional conditions, as already mentioned, are required to establish DI. As mentioned, the Hamiltonians are well behaved. and manageable. We conclude in section 4.\\

\section{Geometry of symmetric teleparallel gravity theory.}

While the Riemann curvature tensor $({R^\alpha}_{\beta\mu\nu})$ based on Levi-Civita connection is the building block of GTR, the general affine connection is a combination of the `Levi-Civita connection', the `Contortion tensor' and the `Disformation tensor' and is expressed as,

\be\label{2.1} {\Gamma^\alpha}_{\mu\nu} = \{_\mu{^\alpha}_\nu\}  + {K^\alpha}_{\mu\nu} + {L^\alpha}_{\mu\nu},\ee
where, the Levi-Civita connection, the contortion tensor and the disformation tensor are given respectively as,
\be\label{2.2} \begin{split}&\{_\mu{^\alpha}_\nu\} = \frac{1}{2} g^{\alpha\lambda}\left(g_{\mu\lambda,\nu}+g_{\nu\lambda,\mu}-g_{\mu\nu,\lambda}\right),\\&
{K^\alpha}_{\mu\nu} = \frac{1}{2}g^{\alpha\lambda}\left(\mathrm {T}_{\lambda\mu\nu}-\mathrm{T}_{\mu\lambda\nu}-\mathrm{T}_{\nu\lambda\mu}\right),\\&{L^\alpha}_{\mu\nu} = \frac{1}{2}g^{\alpha\lambda}\left(Q_{\lambda\mu\nu}-Q_{\mu\lambda\nu}-Q_{\nu\lambda\mu}\right).\end{split}\ee
The Levi-Civita connection is formed out of the derivatives of the metric tensor $(g_{\mu\nu})$, while the contortion and the disformation tensors are formed out of the linear combinations of the torsion tensor $(\mathrm{T}_{\alpha\mu\nu})$ and the non-metricity tensor $(Q_{\alpha\mu\nu})$ respectively. The general Riemann curvature tensor may be expressed in terms of the general affine connection as,
\be \label{Curv} {R^\alpha}_{\beta\mu\nu} = \partial_\mu{\Gamma^\alpha}_{\nu\beta} - \partial_\nu{\Gamma^\alpha}_{\mu\beta} + {\Gamma^\alpha}_{\mu\sigma}{\Gamma^\sigma}_{\nu\beta} - {\Gamma^\alpha}_{\nu\sigma}{\Gamma^\sigma}_{\mu\beta}.\ee
If the connection is not symmetric, the above curvature tensor may be expressed solely out of the contortion tensor, which we denote by ${{\bar R}^\alpha}_{\;\;\beta\mu\nu}$. On the contrary, if the connection is symmetric, it \eqref{Curv} may be expressed in terms of the sum of the Riemann curvature tensor formed out of Levi-Civita connection and a combination of the disformation tensors and we denote it by ${\tilde R^\alpha}_{\;\;\beta\mu\nu}$. In the following, we briefly describe the geometry of the symmetric teleparallel theory.\\

Since generalized gravity theory with torsion $f(\mathrm{T})$ initially had encountered some serious problems with LLI and coupling issues \cite{LLT,fT1l,fT2}, another possible formulation in the flat $({\tilde{R}^\alpha}_{\;\;\beta\mu\nu}  = 0)$ and torsion-free $({\mathrm{T}_{\mu\nu}}^\lambda = 0)$ space-time, often referred to as the symmetric teleparallel theory of gravity, emerged. In symmetric teleparallel gravity, the vanishing curvature constraint imposes the connection to be purely inertial, i.e., it differs from the trivial connection by a general linear gauge transformation. In the absence of torsion, the general affine connection \eqref{2.1} is expressed as ${\Gamma^\alpha}_{\mu\nu} = \{_\mu{^\alpha}_\nu\} + {L^\alpha}_{\mu\nu}$, in which the Christoffel connection mixes inertia with gravitation. As a result, the evolution of the universe is not only controlled by the gravitational effect through ${L^\alpha}_{\mu\nu}$, but also through the inertial effect ${\Gamma^\alpha}_{\mu\nu}$. Inertial effect restores Lorentz covariance and is parametrized by an arbitrary function $\xi^\alpha(x)$, called the St\''uckelberg fields. If connections vanish $({\Gamma^\alpha}_{\mu\nu} = 0)$, as for the coincidence gauge, the inertial effect disappears. Recent focus is on $f(Q)$ theory, i.e., the `generalized symmetric teleparallel gravity' (GSTG) theory. Given a metric tensor  $g_{\mu\nu}$, the only non-trivial object associated to the connection is the non-metricity tensor,
\be\label{2.11}Q_{\alpha\mu\nu} = \nabla_\alpha g_{\mu\nu} = g_{\mu\nu,\alpha} - g_{\nu\rho}{\Gamma^\rho}_{\mu\alpha} -g_{\rho\mu}{\Gamma^\rho}_{\nu\alpha} \ne 0,\ee
where $\nabla$ is the covariant derivative satisfying curvature-free and torsion-free conditions and $Q_{\alpha\mu\nu}$ is symmetric in the last two indices $\mu,\;\nu$. Since the covariant derivative of the metric tensor does not vanish, therefore the term `non-metricity' is used to identify the tensor and the quadratic scalars of symmetric teleparallel gravity. It is possible to construct two different types of non-metricity vectors, viz.,
\be\label{2.12} Q_\alpha= g^{\mu\nu}Q_{\alpha\mu\nu} = {Q_{\alpha\nu}}^\nu;\;\;\;\tilde{Q}_\alpha= g^{\mu\nu}Q_{\mu\alpha\nu} = {Q_{\nu\alpha}}^\nu,\ee
in view of which one can construct a non-metricity conjugate, also called the superpotential tensor as,
\be\label{2.13} P^{\alpha\mu\nu} = -{1\over 4}Q^{\alpha\mu\nu} +{1\over 2}Q^{(\mu\nu)\alpha} +{1\over 4}(Q^\alpha + \tilde{Q}^\alpha)g^{\mu\nu} -{1\over 4}g^{\alpha(\mu}Q^{\nu)},\ee
and finally, the quadratic non-metricity scalar $Q$ expressed as,
\be \label{2.14} Q = -\frac{1}{4}Q_{\alpha\mu\nu}Q^{\alpha\mu\nu} + \frac{1}{2}Q_{\alpha\mu\nu}Q^{\mu\nu\alpha} + \frac{1}{4}Q_\mu Q^\mu - \frac{1}{2}Q_\mu \hat{Q}^\mu = R + \nabla_{\alpha}(Q^{\alpha}-{\tilde{Q}}^\alpha)\ee
is constructed from the curvature-free and torsion-free non-metricity tensor. As mentioned, replacing the curvature scalar by the non-metricity scalar in the EH action \eqref{EH} STEGR emerges. Nonetheless, for an arbitrary functional form $f(Q)$ the action
\be \label{2.15} A_Q = \int f(Q)\sqrt{-g} d^4 x +\mathcal{S}_m,\ee
is called the generalized version of symmetric teleparallel of gravity (GSTG), which deviates from GTR to a great extent. Variation of the above action \eqref{2.15} with respect to the metric tensor $g^{\mu\nu}$ and the connection ${\Gamma^\alpha}_{\mu\nu}$, lead to the following field equations,
\be \label{2.16} \begin{split}&
\frac{2}{\sqrt{-g}} \nabla_\lambda (\sqrt{-g}f_{,Q}{P^\lambda}_{\mu\nu}) +\frac{1}{2}f g_{\mu\nu} + f_{,Q}(P_{\nu\rho\sigma} Q_\mu{}^{\rho\sigma} -2P_{\rho\sigma\mu}Q^{\rho\sigma}{}_\nu) = -\kappa T_{\mu\nu},\\& \nabla_\mu\nabla_\nu\left(\sqrt{-g} f_{,Q} {P^{\mu\nu}}_\lambda\right) = 0,\end{split}\ee
respectively. The metric variation equation \eqref{2.16} may be expressed in the following covariant form \cite{Hoh1,Hoh2},
\be\label{2.16a} f_{,Q} {G}_{\mu\nu} + \frac{1}{2}g_{\mu\nu}\big(Q f_{,Q} - f(Q)\big)+ 2f_{,QQ} {\nabla}_\lambda Q P^\lambda{}_{\mu\nu} = -\kappa T_{\mu\nu},\ee
where, ${G}_{\mu\nu} = {R}_{\mu\nu} - \frac{1}{2} g_{\mu\nu} {R}$ (built from the Levi-Civita connection), and $f_{,Q}$ stands for derivative of $f(Q)$ with respect to $Q$. It is noteworthy that for $f_{QQ} = 0$, the field equation \eqref{2.16a} is simply GTR, while for constant non-metricity scalar $(Q = Q_0)$ it is nothing but GTR with a cosmological constant. In both the cases, the connection variation equation \eqref{2.16a} is trivially satisfied.\\

Now, if the Lie derivative of the connection as well as the space-time metric with respect to the isometry vector ($X$) associated with a space-time metric are set to vanish, i.e.,
\begin{eqnarray}\label{2.17} &\mathcal{L}_X{\Gamma^\mu}_{\alpha\beta} = X^\rho {\partial{\Gamma^\mu}_{\alpha\beta}\over \partial x^\rho} + {\Gamma^\mu}_{\rho\beta}{\partial X^\rho\over \partial x^\alpha} + {\Gamma^\mu}_{\alpha\rho}{\partial X^\rho\over \partial x^\beta} - {\Gamma^\rho}_{\alpha\beta}{\partial X^\mu\over \partial x^\rho} + \frac{\partial^2 X^\mu}{\partial x^\alpha\partial x\beta} = 0,\\&
\mathcal{L}_X {g}_{\mu\nu} = X^\rho \partial_\rho{g}_{\mu\nu} + \partial_\mu X^\rho {g}_{\rho\nu} + \partial_\nu X^\rho {g}_{\mu\rho}=0,
\end{eqnarray}
where, the vector $X^\rho$ is the Killing vector associated with the symmetry of the space-time, then the equations \eqref{2.17} in the curvature-free and torsion-free environment may be utilized to find all possible connections involved with a space-time metric. For curvature-less and torsion-less condition, the affine connection is expressed as\footnote{Under local Lorentz transformation $\xi^\lambda = {\Lambda^\lambda}_\nu x^\nu$, the connection for which curvature tensor vanishes, is expressed in terms of the component of matrix ${\Lambda^\lambda}_\nu$ as ${\Gamma^\alpha}_{\mu\nu} = {(\Lambda^{-1})^\alpha}_\lambda \partial_\mu {\Lambda^\lambda}_\nu$. Further, the torsion vanishes under the condition $\partial_[\mu {\Lambda^\lambda}_{\nu]} = 0$, implying ${\Lambda^\lambda}_\nu = \partial_\nu \xi^\lambda$, where, $\xi^\lambda$ is an arbitrary function of space-time coordinates. Hence, vanishing curvature and torsion condition is given by: ${\Gamma^\alpha}_{\mu\nu} =  {\partial x^\alpha\over\partial \xi^i}\partial_\mu\partial_\nu \xi^i$.},
\be\label{2.18} {\Gamma^\alpha}_{\mu\nu} = {\partial {x^\alpha}\over \partial\xi^\rho}\left({\partial^2 \xi^\rho\over \partial x^\mu\partial x^\nu}\right),\ee
where, the four scalar fields $\xi^\rho (x)^\alpha$ are arbitrary function of $x^\alpha$ and is called the St$\ddot{\mathrm{u}}$eckelberg fields associated with the diffeomorphism.  This transformation ${\Gamma^\alpha}_{\mu\nu}$ preserves the Lorentz covariance in general, under the non-metricity framework. It is noteworthy that under a special choice $\xi^\rho (x) = x^\rho$ (this is possible if the space-time metric can be expressed in cartesian co-ordinate system), the affine connection \eqref{2.18}, being the derivative of the delta function, vanishes. This choice is called the coincidence gauge, for which covariant derivative leads to partial derivative, i.e.,
\be\label{2.19} Q_{\alpha\mu\nu} = \nabla_\alpha g_{\mu\nu} = g_{\mu\nu,\alpha},\ee
and the metric $(g_{\mu\nu})$ becomes the only independent variable. Although, it considerably simplifies all computations and has largely been explored in the literature, nonetheless in the absence of the affine connection the theory is not `Locally Lorentz Invariant' (LLI). This problem may be circumvented, since \eqref{2.18} admits non-trivial connections too. For example in the following isotropic and homogeneous Robertson-Walker metric, which is our current concern,
\be\label{RW} ds^2 = -N^2 dt^2 + a^2(t)\left[\frac{dr^2}{1-kr^2} + r^2(d\theta^2 + r^2 \sin^2 \theta d\phi^2)\right],\ee
where, $N(t)$ is the Lapse function, and $a(t)$ is the scale factor, there exists following six (three translational and three rotational) spatial Killing vectors $X^\rho$
\be \begin{split}& X_1 = \sin{\phi}\partial_\theta + {\cos{\phi}\over \tan{\theta}}\partial_\phi,\;\;X_2 = \cos{\phi}\partial_\theta + {\sin{\phi}\over \tan{\theta}}\partial_\phi,\;\;X_3 = \partial_\phi,\\&
X_4 = \sqrt{1-kr^2}\sin{\theta}\cos{\phi}\partial_r+{\sqrt{1-kr^2}\over r}\cos{\theta}\cos{\phi}\partial_\theta - {\sqrt{1-kr^2}\over r}{\sin{\phi}\over \sin{\theta}}\partial_\phi,\\&X_5 = \sqrt{1-kr^2}\sin{\theta}\sin{\phi}\partial_r+{\sqrt{1-kr^2}\over r}\cos{\theta}\sin{\phi}\partial_\theta - {\sqrt{1-kr^2}\over r}{\cos{\phi}\over \sin{\theta}}\partial_\phi,\\&X_6 = \sqrt{1-kr^2}\cos{\theta}\partial_r+{\sqrt{1-kr^2}\over r}\sin{\theta}\partial_\theta,
\end{split}\ee
which associates non-vanishing affine connections \cite{Hoh1, Hoh2, Qcov2} viz.,
\be\label{Hoh1} \begin{split}& {\Gamma^0}_{00} = K_1,\;{\Gamma^0}_{11} = {K_2\over 1-kr^2},\;{\Gamma^0}_{22} = K_2r^2,\;{\Gamma^0}_{33}=K_2r^2\sin^2 \theta;\\&
{\Gamma^1}_{01} = K_3,\;{\Gamma^1}_{11}= {kr\over 1-kr^2},\;{\Gamma^1}_{22} = -r(1-kr^2),\;{\Gamma^1}_{33} = - r\sin^2{\theta}(1-kr^2);\\&
{\Gamma^2}_{02} = K_3,\;{\Gamma^2}_{12} = {\Gamma^2}_{21} ={1\over r}, {\Gamma^2}_{33} = -\sin{\theta}\cos{\theta};\\&
{\Gamma^3}_{02} = K_3,\;{\Gamma^3}_{13}={\Gamma^3}_{31} = {1\over r},\;{\Gamma^3}_{23} = {\Gamma^3}_{32} = \cot{\theta};\;\;,\end{split}\ee
in view of equation \eqref{2.18}, where $K_i$ are functions of time. There is yet another set of connections possible in the non-flat space, however, as mentioned, in the present article we shall only consider the above three possible connections associated with spatially flat $(k = 0)$ space-time. In the following subsections we study the roll of the connection variation equations in the three spatially flat $(k = 0)$ cases for GSTG gravity theory $f(Q)$ for $f_{QQ} \ne 0$. It may be mentioned that for linear form of $f(Q) \propto Q$, the connection variation equations are trivially satisfied.

\subsection{Connection-1:}
For the first set of connections, henceforth will be called connection-1, $k=0, ~K_1 = \gamma(t),~K_2= K_3 = 0$. Consequently, the non-vanishing components of the affine connections are \eqref{Hoh1},
\be\label{Hoh11} \begin{split}& {\Gamma^0}_{00} = \gamma(t),\;\;{\Gamma^1}_{22} = -r,\;{\Gamma^1}_{33} = - r\sin^2{\theta},\;\; {\Gamma^2}_{12} ={1\over r},\\& {\Gamma^2}_{33} = -\sin{\theta}\cos{\theta},\;\;
{\Gamma^3}_{13} = {1\over r},\;{\Gamma^3}_{23} = \cot{\theta}.\end{split}\ee
As a result, one can compute the non-metricity scalar in view of \eqref{2.14}, which reads as \cite{Dima, Avik},
\be \label{Q1} Q =  -6{\mathcal{H}}^2 =-6{\dot a^2\over N^2 a^2} =-{3\over 2} \left({\dot z^2\over N^2 z^2}\right).\ee
Throughout this article, we treat the Hubble parameter $H = {\dot a\over a}$, while $\mathcal{H}$ stands for  ${\dot a\over Na}$ and the reason for choosing $z = a^2$, will be clarified later. Although, the non-metricity scalar has the same expression as for the coincidence gauge, note that the affine connections are non-vanishing in the spherical coordinate system \eqref{RW} under consideration. Now, the field equations for the metric \eqref{RW} as presented in \cite{Dima,Shi} are given by,
\begin{eqnarray}\begin{split}&\label{FE1}{1\over 2} f + \left(3{\mathcal{H}^2} -{1\over 2}Q\right)f_{,Q} = \rho,\\&
{1\over 2} f + \left(3\mathcal{H}^2 -{1\over 2}Q\right)f_{,Q} + {2\over N}{d\over dt} (\mathcal{H} f_{,Q})= -p,\end{split}\end{eqnarray}
where the matter action $\mathcal{S}_m$ \eqref{2.15} consists of $\rho$ and $p$, which are the energy density and thermodynamic pressure of a barotropic fluid (inclusive of dark matter component) respectively. It is important to note that the second equation of \eqref{2.16}, which is essentially the connection variation equation, is trivially satisfied, since $\gamma(t)$ does not appear in $Q$. Now, let us examine the `energy-momentum conservation law', after setting the cosmic time gauge (lapse function) $N=1$. From the first equation of \eqref{FE1} we find,
\begin{eqnarray}\begin{split} \dot\rho = & \left(3{H^2} -{1\over 2}Q\right)f_{,QQ}{\dot Q}+6{H\dot{H}}f_{,Q},\end{split}\end{eqnarray}
while the combination of first and second equation of \eqref{FE1} results in,
\begin{eqnarray}\label{rp} {\rho+p}=-{d\over dt}(2Hf_{,Q}).\end{eqnarray}
Hence,
\begin{eqnarray} \dot\rho +3H({\rho+p})=& -\left(3{H^2} +{1\over 2}Q\right)f_{,QQ}{\dot Q}=0.\end{eqnarray}
The right hand side of the above equation vanishes in view of the definition of the non-metricity scalar $Q$ and therefore the energy conservation law holds. This is already a known result. Let us repeat that the connection variation equation \eqref{2.16} trivially holds and hence this form of non-metricity scalar administers non-linear form of $f(Q)$ while the field equations are LLI.\\

\noindent
\textbf{Field equations in view of Lagrange multiplier technique:}\\

Lagrange multiplier technique is undebatably a mathematically and physically established technique which has been used frequently to find the field equations of $f(R)$ theory along with its different modified versions, $f(\mathrm{T})$ theory as well as $f(Q)$ theory in coincidence gauge. Although it appears trivial presently, but will turn out to be important for the other two connections, which we shall consider next.\\

Treating the non-metricity scalar as a constraint and introducing it through a Lagrange multiplier $\lambda$, one can recast the action \eqref{2.15} as,
\be\label{A1} A=\int\left[f(Q)- \lambda\left(Q+ {6{\dot a}^2\over N^2a^2}\right) - \rho_0 a^{-\left(3\omega+1\right)}\right]{Na^{3}} dt,\ee
where, a barotropic fluid $p = \omega \rho$, has been inserted, $\omega$ being the state parameter and $\rho_0$ stands for any arbitrary amount of radiation/matter present today, while the constant appearing due to the integration over the three-space is absorbed in the action. Varying the action with respect to $Q$ one gets, $\lambda=f_{,Q}$. Substituting it back in \eqref{A1}, the action finally may be expressed as,
\be\label{A3} A=\int\left[\left\{f(Q)- Q f_{,Q}\right\}a^{3}-{6a{\dot a}^2\over N^2}f_{,Q} - \rho_0 a^{-3\omega}\right] Ndt,\ee
while the field equations are,
\begin{eqnarray}\begin{split}&{1\over 2} f + \left(3{\mathcal{H}^2} -{1\over 2}Q\right)f_{,Q} = \rho,\\&
{1\over 2} f + \left(3\mathcal{H}^2 -{1\over 2}Q\right)f_{,Q} + {2\over N}{d\over dt} (\mathcal{H} f_{,Q})= -p,\\&
Q=-6\mathcal{H}^2,\end{split}\end{eqnarray}
which validates Lagrange multiplier technique.

\subsection{Connection-2:}

Next, we consider the second set of connections (Connection-2) associated with an unknown time dependent parameter $\gamma(t)$. The conditions for spatially flat $(k=0)$ connection-2 are, $K_1 = \gamma(t) + {\dot \gamma\over \gamma};\;K_2 = 0;\;K_3 = \gamma$. Therefore the non-vanishing components of the affine connections are \eqref{Hoh1},
\be\label{Hoh2} \begin{split}& {\Gamma^0}_{00} = \gamma+{\dot\gamma\over\gamma},\; {\Gamma^1}_{01} = \gamma,\;{\Gamma^1}_{22} = -r,\;{\Gamma^1}_{33} = - r\sin^2{\theta};\\&
{\Gamma^2}_{02} = \gamma,\;{\Gamma^2}_{12} ={1\over r}, {\Gamma^2}_{33} = -\sin{\theta}\cos{\theta};\\&
{\Gamma^3}_{02} = \gamma,\;{\Gamma^3}_{13} = {1\over r},\;{\Gamma^3}_{23} = {\Gamma^3}_{32} = \cot{\theta}.\end{split}\ee
Consequently, in view of \eqref{2.14} one finds,
\be\label{Q2} \begin{split}
Q &= -6\mathcal{H}^2 +9\gamma {\mathcal{H}\over N}-3\gamma{\dot N \over N^3} + 3{\dot \gamma\over N^2} \\&=-{6\dot a^2\over N^2 a^2} + {3\gamma\over N^2}\left(3{\dot a\over a} - {\dot N\over N}\right) + {3\dot\gamma\over N^2} = -\frac{3{\dot z}^2}{2z^2N^2}+{3\gamma \over N^2}\left({3\dot z \over 2z}-{\dot N \over N}\right)+{3{\dot \gamma} \over N^2}.\end{split}\ee
The field equations corresponding to connection-2 as presented in \cite{Dima,Shi} are,
\begin{eqnarray}\begin{split}&\label{FE2}{1\over 2} f + \left(3{\mathcal{H}^2} -{1\over 2}Q\right)f_{,Q} + {3\over 2N^2}\gamma \dot Q f_{,QQ} = \rho,\\&
{1\over 2} f + \left(3\mathcal{H}^2 -{1\over 2}Q\right)f_{,Q} + {2\over N}{d\over dt} (\mathcal{H} f_{,Q}) - {3\over 2N^2}\gamma \dot Q f_{,QQ}= -p,\\&
{\dot Q}^2f_{,QQQ}+\left[\ddot Q+\dot Q\left({3\dot a\over a}-{\dot N\over N}\right)\right]f_{,QQ}=0, \Longrightarrow {d\over dt}\left[{a^3\dot Q \over N}f_{,QQ}\right] =0,\end{split}\end{eqnarray}
where the last equation of \eqref{FE2} is an artifact of the second of \eqref{2.16}, which essentially is the connection variation equation. Now, let us first examine the `energy-momentum conservation law', after setting the cosmic time gauge (lapse function) $N=1$. From the first equation of \eqref{FE2} we find,
\begin{eqnarray}\begin{split} \dot\rho = & \left(3{H^2} -{1\over 2}Q\right)f_{,QQ}{\dot Q}+6{H\dot H}f_{,Q}+ {3\over 2}\dot\gamma \dot Q f_{,QQ}+{3\over 2}\gamma {\dot Q}^2 f_{,QQQ}+{3\over 2}\gamma \ddot Q f_{,QQ}.\end{split}\end{eqnarray}
Combination of first and second equation of \eqref{FE2} results in,
\begin{eqnarray}\label{rp1} {\rho+p}=-{d\over dt}(2Hf_{,Q})+3\gamma\dot Q f_{,QQ}.\end{eqnarray}
Hence,
\begin{eqnarray}\begin{split} \dot\rho +3H({\rho+p})=& -\left(3{H^2} +{1\over 2}Q\right)f_{,QQ}{\dot Q}+{3\over 2}\dot\gamma \dot Q f_{,QQ}+{3\over 2}\gamma {\dot Q}^2 f_{,QQQ}+{3\over 2}\gamma \ddot Q f_{,QQ}+9\gamma H\dot Q f_{,QQ}.\end{split}\end{eqnarray}
The `energy-momentum conservation law' holds if and only if the right hand side of the above equation vanishes, while for the connection-1, $(\gamma = 0,~ Q = -6H^2)$, the law holds automatically. But for connection-2, it holds either for $f_{,QQ}=0$ or for $\dot Q=0$, while earlier finding is that, it holds only for $\dot Q = 0$ only \cite{Avik}. From the covariant form \eqref{2.16a}, it is apparent that for $f_{,QQ}=0$ it leads to GTR, whereas for $\dot Q=0$ it represents GTR with a cosmological constant. So, connections-2 cannot be extended beyond GTR or may only be considered in linear scalar-tensor GSTG theory. It is very important to mention that, we have arrived at the above conclusion setting the cosmic gauge $N = 1$, which is true if and only if DI ($\mathbb{H} = N\mathrm{H}$) holds for the Hamiltonian $\mathrm{H}$. Thus the energy-momentum conservation law appears to be an artefact of DI. In the following, we use the the Lagrange multiplier technique to exhibit that the above result $f_{,QQ}=0$ or $\dot Q = 0$, does not require to set $N = 1$.\\

\noindent
\textbf{Field equations in view of Lagrange multiplier technique:}\\

As before, treating the relation for the non-metricity scalar \eqref{Q2} as a constraint and introducing it through a Lagrange multiplier $\lambda$, we recast the action \eqref{2.15} as,
\be \label{AC2}A = \int \left[f(Q) -\lambda\left(Q + {6\dot a^2\over N^2 a^2}-{3\gamma\over N^2}\left(3{\dot a\over a} + {\dot N\over N}\right) - {3\dot \gamma \over N^2}\right) - \rho_0 a^{-3(\omega+1)} \right]Na^3d t.\ee
 Varying the action with respect to $Q$ one gets, $\lambda = {df\over d Q} = f_{,Q} $. Substituting it back, the action \eqref{AC2} may finally be expressed as,
\be\label{A2} A = \int \left[fa^3-Q f_{,Q}a^3-6f_{,Q}{a\dot{a}^2\over N^2} +9\gamma f_{,Q}{a^2\dot{a}\over N^2}-3\gamma a^3f_{,Q}{\dot N\over N^3}+{3f_{,Q}a^3\dot{\gamma}\over N^2} - \rho_0 a^{-3\omega}\right]Nd t.\ee
Now, varying the action with respect to $N, a, Q$ and $\gamma$ respectively, Einstein's equations are obtained as,
\be \label{FEL2}\begin{split} &{1\over 2} f + \left(3{\mathcal{H}^2} -{1\over 2}Q\right)f_{,Q} + {3\over 2N^2}\gamma \dot Q f_{,QQ} = \rho,\\&
{1\over 2} f + \left(3\mathcal{H}^2 -{1\over 2}Q\right)f_{,Q} + {2\over N}{d\over dt} (\mathcal{H} f_{,Q}) - {3\over 2N^2}\gamma \dot Q f_{,QQ}= -p,\\&
Q = -6\mathcal{H}^2 +9\gamma {\mathcal{H}\over N}-3\gamma{\dot N \over N^3} + 3{\dot \gamma\over N^2},\\&
{a^3\dot{Q}\over N}f_{,QQ} = 0.\end{split}\ee
Note that the third, viz., $Q$ variation equation \eqref{FEL2} gives back the definition of $Q$, validating the technique. The most important is the fourth, viz., the $\gamma$ variation (connection variation equation) equation \eqref{FEL2}. It differs from the same that appeared in \eqref{FE2}, since not the derivative, but the expression itself vanishes. This restricts the theory either to a constant non-metricity scalar $Q = Q_{0}$, or to a linear form of the theory $f_{,QQ} = 0$. Neither it is not an outcome of `energy-momentum conservation law', nor the choice of the cosmic time gauge $N$. Thus as mentioned, connections-2 cannot be extended beyond GTR or may only be considered in linear scalar-tensor GSTG theory. \\

\subsection{Connection-3:}
There exists a third set of connections for spatially flat $(k=0)$ case, henceforth will be dubbed as connection-3, under the conditions, $K_1 = - {\dot \gamma\over \gamma};\;K_2 = \gamma;\;K_3 = 0$. The non-vanishing components of the affine connections are \eqref{Hoh1},
\be\label{Hoh2} \begin{split}& {\Gamma^0}_{00} = -{\dot\gamma\over\gamma},\;\;\;\;{\Gamma^0}_{11} = \gamma;\;\;\;\; {\Gamma^0}_{22} = -\gamma r^2,\\&{\Gamma^0}_{33} =  \gamma r^2 \sin^2{\theta},\;{\Gamma^1}_{22} = -r,\;{\Gamma^1}_{33} = - r\sin^2{\theta}.\end{split}\ee
Consequently, in view of \eqref{2.14} the non-metricity  scalar reads as,
\be\label{Q3}\begin{split} Q &= -{6\mathcal{H}^2}+ {3\gamma N\over a^2}\left(\mathcal{H} + {\dot N\over N^2 }\right) + {3\dot\gamma\over a^2}\\& = -6{\dot a^2\over N^2 a^2}+{3\gamma\over a^2}\left({\dot a\over a} + {\dot N\over N}\right) + 3{\dot \gamma \over a^2} = -{3\over 2}{\dot z^2\over N^2 z^2} + {3\gamma\over z}\left({\dot z\over 2z} + {\dot N\over N}\right) + 3{\dot \gamma \over z^2}.\end{split}\ee
Corresponding field equations and the connection variation equation are \cite{Dima,Shi} ,
\begin{eqnarray}\begin{split} \label{FE3a}&{1\over 2} f + \left(3\mathcal{H}^2 -{1\over 2}Q\right)f_{,Q} - 3\gamma{\dot Q f_{,QQ}\over 2a^2} = \rho,\\&
{1\over 2} f + \left(3\mathcal{H}^2 -{1\over 2}Q\right)f_{,Q} +{ 2\over N}{d\over dt} (\mathcal{H} f_{,Q})- \gamma {\dot Q f_{,QQ}\over 2a^2}= -p,\\&
{\dot Q}^2f_{,QQQ}+\left[\ddot Q+\dot Q\left({\dot a\over a}+{\dot N\over N}+{2\dot \gamma\over \gamma}\right)\right]f_{,QQ}=0.\end{split}\end{eqnarray}
Here as before, in order to inspect the `energy-momentum conservation law', we set the cosmic time gauge $(N = 1)$ and find in view of equation \eqref{FE3a},
\begin{eqnarray}\begin{split} \dot\rho =  &\left(3{H^2} -{1\over 2}Q\right)f_{,QQ}{\dot Q}+6{H\dot H}f_{,Q}-{3\over 2a^2}\dot\gamma \dot Q f_{,QQ}-{3\over 2a^2}\gamma {\dot Q}^2 f_{,QQQ}-{3\over 2a^2}\gamma \ddot Q f_{,QQ}+{3\over a^3}\gamma\dot a \dot Q f_{,QQ},\end{split}\end{eqnarray}
while combination of \eqref{FE3a} yields,
\begin{eqnarray}\label{rp2} {\rho+p}=-{d\over dt}(2Hf_{,Q})-\gamma{\dot Q f_{,QQ}\over a^2}.\end{eqnarray}
As a result,
\begin{eqnarray}\begin{split}\dot\rho +3H({\rho+p})= &-\left(3{H^2} +{1\over 2}Q\right)f_{,QQ}{\dot Q}-{3\over 2a^2}\dot\gamma \dot Q f_{,QQ}-{3\over 2a^2}\gamma {\dot Q}^2 f_{,QQQ}-{3\over 2a^2}\gamma \ddot Q f_{,QQ}.\end{split}\end{eqnarray}
Clearly, `the energy-momentum conservation law' holds either for $f_{,QQ}=0$ or for $\dot Q=0$, as before and hence connections-3 also cannot be extended beyond GTR or may only be considered in linear scalar-tensor GSTG theory. Next, we follow Lagrange multiplier method to exhibit, that the above conclusion does not require to set the cosmic time gauge.\\

\noindent
\textbf{Field equations in view of Lagrange multiplier technique:}\\

As before, treating the above connection as a constraint and introducing it through a Lagrange multiplier $\lambda$ into the action and thereafter varying the action as before, the field equations are found as,
\be \label{FE3L}\begin{split}&{1\over 2} f + \left(3\mathcal{H}^2 -{1\over 2}Q\right)f_{,Q} - 3\gamma{\dot Q f_{,QQ}\over 2a^2} = \rho,\\&
{1\over 2} f + \left(3\mathcal{H}^2 -{1\over 2}Q\right)f_{,Q} +{ 2\over N}{d\over dt} (\mathcal{H} f_{,Q}) - \gamma {\dot Q f_{,QQ}\over 2a^2}= -p,\\&
Q = -{6\mathcal{H}^2}+ {3\gamma N\over a^2}\left(\mathcal{H} + {\dot N\over N^2 }\right) + {3\dot\gamma\over a^2},\\&
{a^3\dot{Q}\over N}f_{,QQ}=0.\end{split}\ee
Here again, the $Q$ variation equation  of \eqref{FE3L} gives back the definition of $Q$ validating the technique. However, the $\gamma$ variation equation of \eqref{FE3L} again implies that either $Q = Q_0$, a constant or a linear form of $f(Q)$ is admissible and there is no need to consider the energy-momentum conservation law, or to set the cosmic time gauge $(N)$, either. Thus, connection-3 also cannot be extended beyond GTR, although may be considered in linear scalar-tensor GSTG theory.\\

\subsection{Linear scalar-tensor GSTG theory:}

In a nutshell, it is revealed that connection-1 is only possible to apply in $f(Q)$ GSTG theory for $f_{QQ} \ne 0$. On the contrary, leaving GTR apart, connections 2 and 3 may only be applied in linear GSTG theory with scalar coupling. Now, in conjugation with the scalar-tensor equivalent theory of $f(R)$ gravity, it is also possible to construct a scalar-tensor theory of GSTG in the following form,
\be\label{Action} A_{[Q,\Phi]} = \int [f'(\Phi) Q + f(\Phi) - \Phi f'(\Phi)]\sqrt{-g} d^4x + \mathcal{S}_m,\ee
where, prime now denotes derivative with respect to the dynamical auxiliary variable $\Phi$. Unfortunately, the above action \eqref{Action} is \eqref{2.15} in disguise, which is apparent taking the variation of action \eqref{Action} with respect to $\Phi$, whence one finds $\Phi = Q$. Therefore action \eqref{Action} is the standard LLI $A_{Q,\Gamma}$ theory for which only connection-1 is applicable. Nonetheless, the auxiliary field $\Phi$ introduced for linearizing the theory turns out to be a ghost (negative norm state) in the presence of the shift vector. However, there is yet another technique that generalizes STEGR, namely the scalar-vector-tensor formalism \cite{Qcov1}, in accordance to the familiar scalar-curvature (scalar-tensor) theory. The symmetric teleparallel action in the very presence of the term $(Q^\alpha - \hat{Q}^\alpha)\partial_\alpha f(\Phi)$ may be cast in the following form \cite{Qcov1, Hu},
\be \label{Qcov1} A_{\mathrm{Cov}[Q,\Gamma]} = A_{\mathrm{Cov}} = \int \left[{1\over 2}\{f(\Phi)Q + \left(Q^\alpha - \hat{Q}^\alpha\right)f_{,\alpha}(\Phi)\} -\omega(\Phi)\Phi_{,\mu}\Phi^{,\mu} - U(\Phi)-\phi_{,\alpha} \phi^{,\alpha}-V(\phi)\right]\sqrt{-g} d^4 x +  \mathcal{S}_m.\ee
The very presence of the second term makes the action \eqref{Qcov1} covariant under the conformal transformations and scalar field redefinitions. Further, such a linear theory is free from extra degrees of freedom and hence ghosts do not appear. We therefore shall formulate the phase-space structure for the action \eqref{2.15} with connection-1, and the same for the scalar-vector-tensor action \eqref{Qcov1} taking into account all the three connections. It may be mentioned that the problem with extra degrees of freedom and consequently the strong coupling issue along with the ghost-degrees of freedom do not appear in the absence of the shift vector $(N^i)$, as we shall explain later.

\section{Hamiltonian formulation and Diffeomorphic invariance:}

In the present manuscript, our aim is to formulate the phase-space structure in the spatially flat RW metric, starting from the non-LLI scalar-non-metricity action \eqref{Action}, for the sake of comparison with the LLI cases. This requires to perform Dirac-Bergmann constraint analysis. Next, we shall consider the LLI scalar-vector-tensor action \eqref{Qcov1} and consider the non-metricity scalar $Q$ constructed out of non-vanishing affine connections. Let us recall that all the earlier works in this direction, although were associated with a general space-time metric, nevertheless were performed with non-LLI actions and assuming coincidence gauge. ADM $(3+1)$ decomposition and Hamiltonian formulation of STEGR $[f(Q) \propto Q]$ was performed in the general space-time background in \cite{H1}. The result is trivial since both the lapse function $(N)$ and the shift vector $(N^i)$ act as dynamical variables in the linear case. The Hamiltonian formulation of $f(Q)$ gravity in coincident gauge, has been executed by several authors following Dirac-Bergmann constraint analysis, starting from the action \eqref{Action}, which is not Lorentz invariant \cite{H2,H3,H4}. These analysis resulted in additional degrees of freedom and the presence of ghosts. To be precise, the presence of non-propagating variables such as the shift vector $(N^i)$ and the auxiliary variable $\Phi$ resulted in the violation of  diffeomorphic invariance (DI) \cite{H2}. Authors \cite{H3} on the contrary, proved that Dirac-Bergmann algorithm itself fails, since the consistency condition appears with an integration, which requires to solve a set (instead of one) of inhomogeneous system of partial differential equations, that cannot be solved exactly. In a recent work, the authors \cite{H4} reviewed the issue raised in \cite{H3} extensively and inferred that the problem may be circumvented by imposing appropriate spatial boundary conditions, although DI is violated. In another work, disappearance of ghost has been claimed in the conformally transformed covariant scalar-vector-tensor of GSTG, nonetheless, neither DI has not been checked nor the Hamiltonian was formulated \cite{Hu}. In a nut-shell, if the set of eleven variables $(N,N^i,h_{ij},\Phi)$ are considered, then DI cannot be established. The trouble stem from the presence of shift vector $(N^i)$ and the auxiliary variable $\Phi$, which are found to be the dynamical non-propagating variables carrying $1\over 2$ degree of freedom each. A closer look to these works reveal that, the momentum associated with the auxiliary variable $\Phi$ appears as the first derivative of the shift vector, $p_\Phi = {\sqrt h\over N}f''(\partial_i N^i)$. Thus, in the absence of the shift vector, i.e., if a minisuperspace model of cosmological interest being devoid of the shift vector is taken into account, then the momentum $p_{N^i}$ does not appear and as a result $p_\Phi$ would constraint to vanish. In that case, non-LLI action may establish DI unambiguously and consequently the structure of the Hamiltonian would be revealed. Therefore, we choose a minisuperspace model of cosmological interest such as the Robertson-Walker metric \eqref{RW}, to get rid of the shift vector. This finally ends up with a specific form of the Hamiltonian, that would throw some light in regard of studying inflation in view of the classical field equations. However, all these works along with the said one cannot be taken seriously, due to the choice of the coincidence gauge. In this manuscript we consider GSTG actions \eqref{Action} as well as \eqref{Qcov1}, but consider non-trivial connections to explore DI and construct the Hamiltonian. It is found that the issue of DI in the absence of the shift vector might not be trivial.\\

It is customary to express the Hamiltonian in terms of the scale factor `$a(t)$' although the induced three metric $h_{ij} = a^2 \delta_{ij} = z \delta_{ij}$ is the true basic variable upon $(3+1)$ ADM decomposition. In fact, no problem appears ordinarily, unless Hamiltonian formulation of higher-order theory of gravity, such as f(R), is invoked, in which the `induced three metric' and the `extrinsic curvature tensor' $\{h_{ij},~K_{ij}\}$ are to be treated as basic variables. The reason being, these variable can only remove the appropriate surface terms (which are found under variation of the action) upon integration by parts. On the contrary, if one starts with the scale factor $(a)$ in the case of $R^2$ gravity for example, a redundant term such as ${2\over 3}\dot a^3$ is eliminated from the action upon integration by parts \cite{1}. This results in a completely different and uncanny quantum equation. \\

The main objective of the present work is therefore to re-explore earlier works \cite{H2,H3,H4} by considering the actions \eqref{Action} and \eqref{Qcov1} in a minisuperspace of cosmological interest, being devoid of the shift vector, such as the isotropic and homogeneous Robertson-Walker metric \eqref{RW}. To handle action \eqref{Qcov1}, we need to compute the term $\left(Q^\alpha-\hat{Q}^\alpha\right)\partial_\alpha f(\Phi)$, which is essentially $\left(Q^0-\hat{Q}^0\right)f'(\Phi)\dot\Phi$, since $\Phi = \Phi(t)$. The required non vanishing components of non-metricity tensor and the corresponding vectors may be computed in view of \eqref{2.11} and \eqref{2.12} respectively as,
\be\label{TVC}\begin{split}& Q_{000} = -2N^3\left({\dot N\over N^2} - {K_1\over N}\right);\\&
{Q^{0i}}_i = -{2\over N^2}\left(K_3 - H\right);\;\;\; {Q^{i0}}_i = {1\over N^2}\left(-K_3 + {K_2\over a^2}\right);\;\;\; {Q^{00}}_0 = -{2\over N}\left({\dot N\over N^2} - {K_1\over N}\right).\end{split}\ee
Consequently, one finds,
\be\label{cterm} \left(Q^\alpha-\hat{Q}^\alpha\right)\partial_\alpha f(\Phi) = \left(Q^0-\hat{Q}^0\right)f'(\Phi)\dot\Phi = {3\over N^2}\left(3K_3 - 2H \right)- {3K_2\over a^2},\ee
where $f'(\Phi) = f_{,\Phi}$, i.e., derivative with respect to $\Phi$. Correspondingly, one can finally compute the non-metricity scalar $Q$ in view of \eqref{2.14} as well.

\subsection{The phase-space structure for action \eqref{Action} (Connection-1):}

In view of \eqref{2.14}, the scalar-tensor action \eqref{Action} of GSTG may be expressed as,
\be\label{phiA1} A=\int\left[ f'(\Phi)\left\{R + \nabla_{\alpha}(Q^{\alpha}-{\tilde{Q}}^\alpha)\right\}+f(\Phi)-\Phi f'(\Phi)+L_m \right]\sqrt{-g} d^4x.\ee
ADM formulation of the action goes as follows. Under $(3+1)$ decomposition
\be g_{\mu\nu} =
 \left(
  \begin{array}{cc}
    N^2+N_i N^i & N^i \\
    N_i & h_{ij}\\
  \end{array}
\right)\ee
the curvature scalars are expressed in terms of the extrinsic curvature tensor $(K_{ij})$ and its trace $(K)$ as,
\be {^4R} = {^3R} + K_{ij}K^{ij} - K^2,\hspace{3 mm}{^3R} = {^3Q} -D_l(^3Q^l - {^3\tilde{Q}}^l)\ee
and the action \eqref{phiA1} may finally be expressed as \cite{H2}
\be\label{phiA2}\begin{split}& A = \int\Big[f(\Phi) + f'(\Phi)\left({^3Q}+K_{ij}K^{ij} - K^2-\Phi\right) -D_l \{f'(\Phi)(^3Q^l - {^3\tilde{Q}}^l)\} \\&\hspace{2 cm} + {1\over N^2}\left(\dot\Phi f''(\Phi) \partial_i N^i - \dot{N}^i\partial_i f'(\Phi) - \partial_i\{ f'(\Phi)(N^i\partial_j N^j - N^k\partial_k N^i)\}\right)\Big]N{\sqrt h} d^4x + \mathcal{S}_m.\end{split}\ee
Now in the absence of the shift vector, as in the RW metric \eqref{RW}, the action \eqref{phiA2} is drastically simplified to,
\be\label{phiA3} A = \int\left[f(\Phi) + f'(\Phi)\left({^3Q}+K_{ij}K^{ij} - K^2-\Phi\right) -D_l \{f'(\Phi)(^3Q^l - {^3\tilde{Q}}^l)\}\right]N{\sqrt h} d^4x + \mathcal{S}_m.\ee
Nonetheless for the present purpose, it suffices to consider the action \eqref{phiA1}, which has been validated for connection-1 only. In terms of the function $z = a^2$, the action \eqref{phiA1} reads as,
\be\begin{split}\label{phiA2.2} A&=\int\left[\left\{f(\Phi)- \Phi f'(\Phi)\right\}{Nz^{3\over 2}}-{3{\dot z}^2\over 2{N\sqrt z}}f'(\Phi)+\left({1\over 2N^2} \dot\phi^2 - V(\phi)\right) Nz^{3\over 2}\right]dt,\end{split}\ee
where, we introduce the scalar field $\phi$ which would drive inflation. In view of the above equation \eqref{phiA2.2}, the canonical momenta are found as,
\be \label{Ap1} p_{z}={\partial L\over \partial{\dot z}}=-{3{\dot z}\over {N\sqrt z}}f'(\Phi),\hspace{0.2 in} p_{\phi}={\partial L\over \partial{\dot \phi}}={\dot\phi z^{3\over 2}\over N}, \hspace{0.2 in} p_{\Phi}={\partial L\over \partial{\dot \Phi}} = 0, \hspace{0.2 in} p_{N}={\partial L\over \partial{\dot N}} = 0.\ee
Therefore, one encounters two primary constraints associated with the theory, viz.,
\be\label{Aconstraints}\lambda_1 = p_{\Phi} \approx 0, \hspace{0.2 in}\lambda_2 = p_N \approx 0,\ee
which are second class. However note that, because none of the momenta densities contain $\dot N$, so the correlated constraint vanishes strongly. Therefore, the lapse function is non-dynamical and it may be safely ignored. By employing the same argument, one might prompt to ignore $\Phi$ as well. However, it has been shown that in the conformal transformed action, $\dot \sigma$ appears, which is the conformal field that replaces $\Phi$ \cite{Hu}. Therefore, it is not legitimate to ignore $\Phi$, although we do not work in the conformally transformed frame. Note that treating $\Phi$ as non-dynamical, a simple form of the Hamiltonian was obtained in \cite{Cap}, which is not correct. Here therefore, we tacitly assume that the auxiliary field is dynamical. Hence, it is only required to introduce the constraint associated with $p_{\Phi}$ through Lagrange multipliers $u_{1}$ into the following constrained Hamiltonian as,
\be \label{AHC} \mathbb{H}_c =\sum_i p_i \dot q_i - L = \dot z p_z + \dot \phi p_\phi +\dot \Phi p_\Phi+ \dot N p_N - L,\ee
and the primary hamiltonian may be expressed as,
\be\label{AHp}\begin{split} \mathbb{H}_{p} &= \mathbb{H}_c + u_1 p_{\Phi}
=-\frac{{p_z}^2N\sqrt z}{3f'}+\frac{3{\dot z}^2f'}{2N\sqrt z}-\left(f-\Phi f'\right)Nz^{3\over 2}+\frac{Np_{\phi}^2}{2z^{3\over 2}}+ NV(\phi)z^{3\over 2}+ u_1 p_\Phi \\& = -\frac{{p_z}^2N\sqrt z}{6f'}-\left(f-\Phi f'\right)Nz^{3\over 2}+\frac{Np_{\phi}^2}{2z^{3\over 2}}+ NV(\phi)z^{3\over 2}+ u_1 p_\Phi. \end{split}\ee
Note that the Poisson bracket $\{z,p_{z}\}=\{\phi, p_{\phi}\} = \{\Phi,p_{\Phi}\}=1$ holds. Now, constraint should remain preserved in time, which is exhibited through the following poisson's bracket,
\be{\dot\lambda_{1}}=\left\{ \lambda_{1}, \mathbb{H}_{p}\right\}= -Nf''\left[\Phi z^{3\over 2}+\frac{{p_z}^2\sqrt z}{6f'^2}\right]+\sum^2_{i=1}\lambda_i\{\lambda_1,u_i\}\approx 0.\ee
Hence, the Lagrange multiplier $u_1$ remains obscure and the primary constraint $\lambda_1$ results in a secondary constraint, viz.,
\be \label{Aconst2}\lambda_2= -{Nf''}\left[\Phi z^{3\over 2}+\frac{{p_z}^2\sqrt z}{6f'^2}\right]\approx 0,\ee
The above constraint is required to be introduced into the primary Hamiltonian \eqref{AHp}, through yet another Lagrange multiplier. The modified primary Hamiltonian therefore reads as,
\be\label{AHc1} \mathbb{H}_{p1}= -\left(f-\Phi f'\right)Nz^{3\over 2}-\frac{N{p_z}^2\sqrt z}{6f'}+\frac{Np_{\phi}^2}{2z^{3\over 2}}+ NV(\phi)z^{3\over 2}+ u_{1}p_{\Phi}-u_2Nf''\left[\Phi z^{3\over 2}+\frac{{p_z}^2\sqrt z}{6f'^2}\right],\ee
where $u_2$ is yet another Lagrange multiplier. However at this end to preserve the constraints over time, we compute the Poisson brackets yet again, with the modified primary Hamiltonian to find,
\be\label{APoisson1}\begin{split}&{\dot\lambda_{1}}=\left\{ \lambda_{1}, \mathbb{H}_{p1}\right\}=-Nf''\left[\Phi z^{3\over 2}+\frac{{p_z}^2\sqrt z}{6f'^2}\right]+ u_2 N\left[f''\big(z^{3\over 2}-\frac{{p_z}^2\sqrt z f''}{3f'^3}\big)+f'''\big(\Phi z^{3\over 2}+\frac{{p_z}^2\sqrt
z}{6f'^2}\big)\right]\\&\hspace{5.0 cm}+\sum^2_{i=1}\lambda_i\{\lambda_1,u_i\}\approx 0,
\\&{\dot\lambda_{2}}=\left\{ \lambda_{2}, \mathbb{H}_{p1}\right\}=N^2f''\left(\frac{\Phi z{p_z}}{f'}-\frac{z{p_z}f}{2f'^2}- \frac{z p_z p_\phi^2}{4f'^2 z^{5\over 2}} +\frac{z p_z V(\phi}{2f'^2}\right)\\& \hspace{2.0 cm}-u_1N{\left[f''\big(z^{3\over 2}-\frac{{p_z}^2{\sqrt z}f''}{3f'^3}\big)+f'''\big(\Phi z^{3\over 2}+\frac{{p_z}^2\sqrt z }{6f'^2}\big)\right]}+\sum^2_{i=1}\lambda_i\{\lambda_2,u_i\}\approx 0.\end{split}\ee
The above Poisson brackets reveals the following forms of the two Lagrange multipliers $u_1$ and $u_2$,
\be\label{ALagmult2}\begin{split}& u_1=\frac{Nf'' z p_z\left(\frac{\Phi}{f'}-\frac{f}{2f'^2} - \frac{p_\phi^2}{4f'^2 z^{5\over 2}} +\frac{ V(\phi)}{2f'^2}\right)}{\left[f''\big(z^{3\over 2}-\frac{{p_z}^2{\sqrt z}f''}{3f'^3}\big)+f'''\big(\Phi z^{3\over 2}+\frac{{p_z}^2\sqrt z }{6f'^2}\big)\right]} \\&u_2=\frac{f''\left(\Phi z^{3\over 2}+\frac{{p_z}^2\sqrt z }{6f'^2}\right)}{\left[f''\big(z^{3\over 2}-\frac{{p_z}^2\sqrt z f''}{3f'^3}\big)+f'''\big(\Phi z^{3\over 2}+\frac{{p_z}^2\sqrt z }{6f'^2}\big)\right]}.\end{split}\ee
Now, replacing the auxiliary variable $\Phi$ by $Q$, so that $f = f(Q)$, while prime now denotes derivative with respect to $Q$; and using the definition of $p_z$ \eqref{Ap1}, one finds
\be f''\left[Q z^{3\over 2} + \frac{9\dot z^2}{N^2 z} \times f'^2\left({\sqrt z\over 6 f'^2}\right)\right] = f''z^{3\over 2}\left[Q + {3\over 2}\left({\dot z^2\over N^2 z^2}\right) \right] = 0,\ee
using the definition of $Q$. Clearly, the Lagrange multiplier $u_1$ is simplified while the Lagrange multiplier $u_2$ vanishes identically. Hence, one finally ends up with,
\be\label{Lagmult3}u_1 = \frac{\left(\frac{Q}{f'}-\frac{f}{2f'^2} - \frac{p_\phi^2}{4f'^2 z^{5\over 2}} +\frac{ V(\phi)}{2f'^2}\right)}{\big(z-\frac{{p_z}^2 f''}{3f'^3}\big)} N \sqrt{z} p_z = \frac{\left(Q f'-\frac{f}{2} - \frac{p_\phi^2}{4 z^{5\over 2}} +\frac{ V(\phi)}{2}\right)}{\big(zf'^3-\frac{{p_z}^2 f''}{3}\big)} N f'\sqrt{z} p_z,\ee
remembering the fact that for generalized teleparallel gravity theories $f'' \ne 0$. In view of the above form of Lagrange multiplier $u_1$ \eqref{Lagmult3}, the Hamiltonian being free from the constraints may finally be expressed as,
\be\begin{split} \label{H1} \mathbb{H}= &N\left[-\left(f-Q f'\right)z^{3\over 2}-\frac{{p_z}^2\sqrt z}{6f'}+\frac{p_{\phi}^2}{2z^{3\over 2}}+ V(\phi)z^{3\over 2}+ \frac{\left(Q f'-\frac{f}{2} - \frac{p_\phi^2}{4 z^{5\over 2}} +\frac{ V(\phi)}{2}\right)}{\big(zf'^3-\frac{{p_z}^2 f''}{3}\big)} f'\sqrt{z}~ p_{Q}p_z\right]= N \mathrm{H},\end{split}\ee
where the Hamiltonian is,
\be \label{H22} \mathrm{H} = -\left(f- Qf'\right)z^{3\over 2}-\frac{{p_z}^2\sqrt z}{6f'}+\frac{p_{\phi}^2}{2z^{3\over 2}}+ V(\phi)z^{3\over 2}+\frac{\left(Q f'-\frac{f}{2} - \frac{p_\phi^2}{4 z^{5\over 2}} +\frac{ V(\phi)}{2}\right)}{\big(zf'^3-\frac{{p_z}^2 f''}{3}\big)} f'\sqrt{z}~p_{Q}p_z.\ee
Diffeomorphic invariance is thus established and there exists no ghost degree of freedom. It is no surprise, since the root of trouble originated from the shift vector. However, although a cosmological model of interest appears to be trouble free, alas, the Hamiltonian looks quite formidable due to the presence of $p_z^2$ term in the denominator, as obtained earlier while analyzing for GMTG theory \cite{MKA}. More importantly, the minimally coupled scalar field has now been coupled with $f'(Q)$ in the Hamiltonian. Therefore, the Hamiltonian is not only eerie, but also quite impossible to handle as such. Note that the Hamiltonian is tractable only if ${p_z}^2 f''$ is a constant, which results in $z Q f'^2f'' = c$, where $c$ is a constant, but in that case too, one can not avoid unethical coupling between the scalar field $\phi$ and $f'(Q)$. In any case, since $z$ cannot be expressed in terms of $Q$, the form of $f(Q)$ also remains obscure. Although it is possible to canonically quantize the above Hamiltonian \eqref{H22}, however the presence of fourth degree in momenta makes it unhealthy and there is no way to set up a semiclassical formulation. Inflation has been studied mostly choosing a specific form of $f(Q) = \alpha Q + \beta Q^2$, which in no way simplifies the form of the above Hamiltonian \eqref{H22}.\\

Summarily, in the RW minisuperspace model, GSTG [$f(Q), ~f_{QQ} \ne 0$] theory admits DI. Unfortunately, the Hamiltonian is not well-behaved. The modified Wheeler-DeWitt equation although may be established, it is impossible to compute semi-classical approximation and to check `Hartle criteria'. The study of inflation in view of the classical field equations without a viable semiclassical formulation has no merit. In this respect, $f(Q), ~f_{QQ} \ne 0$ theory lacks credential.

\subsection{The phase-space structure for action \eqref{Qcov1} ~(Connection-1):}

Here, we start with the scalar-vector-tensor form of covariant action \eqref{Qcov1} and use connection-1. We first compute
\be\label{Qa1}\left(Q^\alpha - \hat{Q}^\alpha\right)f_{,\alpha}(\Phi) = - {6H\over N^2}f'({\Phi})\dot \Phi = -\frac{3\dot z}{N^2z}f'{\dot\Phi},\ee
in view of \eqref{cterm}. Next, using equations \eqref{Q1} and \eqref{Qa1}, the action \eqref{Qcov1} is expressed as,
\be\begin{split} \label{A2} A_{\mathrm{cov}1}&=\int\left[ -{3{\dot z}^2\over 4{N\sqrt z}}f(\Phi)-\frac{3{\dot z\sqrt z}f'{\dot\Phi}}{2N}+\left({1\over N^2}\omega(\Phi) {\dot\Phi}^2 - U(\Phi)\right) Nz^{3\over 2}+\left({1\over 2N^2} {\dot\phi}^2 - V(\phi)\right) Nz^{3\over 2}\right]dt,\end{split}\ee
where, $\phi$ is the additional scalar field required to drive inflation and $U(\Phi)$, $V(\phi)$ are the potentials associated with $\Phi$ and $\phi$ respectively. The point lagrangian therefore reads as,
\be \label{Cov1}L_{\mathrm{cov}1} = -{3{\dot z}^2\over 4{N\sqrt z}}f(\Phi)-\frac{3{\dot z\sqrt z}f'{\dot\Phi}}{2N}+\left({1\over N^2}\omega(\Phi) {\dot\Phi}^2 - U(\Phi)\right) Nz^{3\over 2}+\left({1\over 2N^2} {\dot\phi}^2 - V(\phi)\right) Nz^{3\over 2}. \ee
The corresponding canonical momenta are,
\be\label{Ap} p_{z}=-{3{\dot z}\over {2N\sqrt z}}f(\Phi)-{3{\sqrt z}\over {2N}}f'(\Phi)\dot\Phi,\hspace{0.2 in} p_{\phi}={\dot\phi z^{3\over 2}\over N}, \hspace{0.2 in} p_{\Phi} ={2\omega(\Phi) z^{3\over 2}{\dot\Phi}\over N} , \hspace{0.2 in} p_{N} = 0\ee
Clearly, one encounters one primary constraint associated with the theory, viz.,
\be\label{Aconstraints}\lambda_1 = p_N \approx 0,\ee
which is a second class constraint and the Poisson bracket $\{z,p_{z}\}=\{\phi, p_{\phi}\} = \{\Phi,p_{\Phi}\}=\{N,p_N\}=1$ holds. However, since none of the momentum densities contain $\dot N$, so the correlated constraint vanishes strongly. We can therefore safely neglect the constraint \eqref{Aconstraints} and express the Hamiltonian in the following form,
\be \begin{split}\label{AHc} \mathbb{H}_c =&\sum_i p_i \dot q_i - L = \dot z p_z + \dot \phi p_\phi +\dot \Phi p_\Phi+ \dot N p_N - L\\&
=-{3{\dot z}^2\over 4{N\sqrt z}}f(\Phi)+\left({1\over N^2}\omega(\Phi) {\dot\Phi}^2 + U(\Phi)\right) Nz^{3\over 2}+\left({1\over 2N^2} {\dot\phi}^2 + V(\phi)\right) Nz^{3\over 2},\end{split}\ee
and consequently,
\be \label{AHC}\mathbb{H}_{\mathrm{Cov1}}=N\left[-\frac{{\sqrt z} p_z^2}{3f}+\left(1-{3f'^2\over 4\omega f}\right)\frac{p_{\Phi}^2}{4\omega z^{3\over 2}}+\frac{p_{\phi}^2}{2z^{3\over 2}}-\frac{z f'p_z p_{\Phi}}{2f\omega z^{3\over 2}}+U(\Phi)z^{3\over 2}+V(\phi)z^{3\over 2}\right] = N\mathrm{H}.\ee
Clearly, not only that DI is established again, but also the Hamiltonian being quadratic in momenta is pertinent and quite simple in structure. Upon appropriate operator ordering between $\{z,p_z\}$, and $\{\Phi,p_\Phi\},$ the Hamiltonian leads to the modified Wheeler-De Witt equation when quantization. Thus, semiclassical approximation is standard and Hartle condition is met once the form of $f'(\Phi)$ is known by some means, say upon requiring a vacuum de-Sitter solution.

\subsection{The phase-space structure for action \eqref{Qcov1}~ (Connection-2):}

In view of \eqref{cterm} one can compute,
\be\label{Qa2} \left(Q^{\alpha}-{\tilde{Q}}^\alpha\right) f_{,\alpha}(\Phi)={3\over N^2}\big(3\gamma-2H\big)f'{\dot\Phi}={3\over N^2}\left(3\gamma-{\dot z\over z}\right)f'{\dot\Phi}.\ee
Substituting equations \eqref{Q2} and \eqref{Qa2}, the action \eqref{Qcov1} now reads as,
\be\begin{split} \label{A2} A_{\mathrm{cov}2} &=\int\Bigg[ -{3{\dot z}^2\over 4{N\sqrt z}}f(\Phi)+\frac{3\gamma f}{2N}\left({3\over 2}\dot z \sqrt z- \frac{\dot N z^{3\over 2}}{N}\right)+\frac{3\dot \gamma fz^{3\over 2}}{2N}+{3\over 2N}\left(3\gamma z^{3\over 2}-\sqrt z\dot z\right)f'{\dot \Phi}\\&\hspace{4.0 cm}+\left({1\over N^2}\omega(\Phi) {\dot\Phi}^2 - U(\Phi)\right) Nz^{3\over 2}+\left({1\over 2N^2} {\dot\phi}^2 - V(\phi)\right) Nz^{3\over 2}\Bigg]dt.\end{split}\ee
The canonical momenta are:
\be\begin{split}& p_z = - {3\dot z f\over 2N\sqrt z} + {9\sqrt {z} \gamma f\over 4N} - {3\sqrt z f'\dot\Phi\over 2N};\;\;p_\Phi =  {9\gamma z^{3\over 2}f'\over 2N} - {3\sqrt{z}\dot z f'\over 2N} + {2\omega z^{3\over 2}\dot \Phi\over N};\\&
p_\gamma = {3 z^{3\over 2}f\over 2N};\;\;\;\; p_N = -{3\gamma z^{3\over 2}f\over 2N^2},~~~p_\phi ={\dot\phi z^{3\over 2}\over N}.\end{split}\ee
Thus, one may compute $\dot z$ and $\dot\Phi$ from the expressions of  $p_z$ and $p_{\Phi}$ respectively as,
\be\begin{split}&{\dot z} =-{2N\sqrt{z}\over 3f}p_z + {3\over 2}\gamma z - {z f'{\dot\Phi}\over f}
= -{2N\sqrt{z}\over 3f}p_z + {3\over 2}\gamma z - {2N f'\over \sqrt{z} f}\left[{f p_\Phi - z f' p_z - {9\over 4N}\gamma z^{3\over 2}f f'\over 3f'^2 + 4 \omega f}\right],\\&
{\dot\Phi}={2N\over z^{3\over 2}}\left[{f p_\Phi - z f' p_z - {9\over 4N}\gamma z^{3\over 2}f f'\over 3f'^2 + 4 \omega f}\right].\end{split}\ee
However, there exists a second class constraint $\lambda_1 = Np_N + \gamma p_\gamma \approx 0$ which is required to introduce into the constrained Hamiltonian through a Lagrange multiplier $(u_1)$, to cast the primary Hamiltonian,
\be\begin{split}\label{PH} \mathbb{H}_{p2} =& \mathbb{H}_c + u_1(Np_N + \gamma p _\gamma) = \dot z p_z + \dot \Phi p_\Phi + \dot N p_N + \dot \gamma p_\gamma + \dot \phi p_\phi - L + u_1(Np_N + \gamma p _\gamma)\\&=-{N\sqrt{z}\over 3f}p_z^2 + {3\gamma\over 2} z p_z + {Np_\phi^2\over 2 z^{3\over 2}} - {27\over 16N} \gamma^2 z^{3\over 2}f + N(U + V)z^{3\over 2}\\&
+ {N\over 3f'^2 + 4 \omega f}\left({f'^2\over f} \sqrt{z} p_z^2 + {f \over z^{3\over 2}}p_\Phi^2 - {2f'\over \sqrt{z}}p_z p_\Phi\right)\\&+ {1\over 3f'^2 + 4 \omega f}\left({9\over 2}\gamma f'^2 z p_z - {9\over 2}\gamma f f' p_\Phi + {81\over 16N}\gamma^2 z^{3\over 2} ff'^2\right)+u_1(Np_N + \gamma p _\gamma).\end{split}\ee
At this end to preserve the constraints over time, we compute the Poisson bracket, with the primary Hamiltonian to find,
\be\label{APoisson12}\begin{split}{\dot\lambda_{1}}=\left\{ \lambda_{1}, \mathbb{H}_{p2}\right\}= &-N\Bigg[-{\sqrt{z}\over 3f}p_z^2 + {p_\phi^2\over 2 z^{3\over 2}} + {27\over 16N^2} \gamma^2 z^{3\over 2}f + (U + V)z^{3\over 2}+{1\over 3f'^2 + 4 \omega f}\bigg({f'^2\over f} \sqrt{z} p_z^2 + {f \over z^{3\over 2}}p_\Phi^2 \\&- {2f'\over \sqrt{z}}p_z p_\Phi-\frac{81\gamma^2ff'^2z^{3\over 2}}{16N^2}\bigg)\Bigg]-\gamma\Bigg[{3\over 2} z p_z-{27\over 8N} \gamma z^{3\over 2}f\\&-{\gamma\over {3f'^2 + 4 \omega f}}\bigg({9\over 2} f'^2 z p_z - {9\over 2}f f' p_\Phi + {81\over 8N}\gamma z^{3\over 2} ff'^2\bigg)\Bigg]\approx 0\end{split},\ee
which is again a second class constraint.\\

\noindent
\textbf{Inspection:}\\
The primary Hamiltonian is found to contain terms such as, $\gamma zp_z, {\gamma^2 fz^{3\over 2}\over N}, {\gamma f'^2 z p_z \over  3f'^2 + 4 \omega f}, {\gamma ff' p_\Phi\over 3f'^2 + 4 \omega f}, {\gamma^2 ff'^2 z^{3\over 2}\over 3f'^2 + 4 \omega f}$, which are devoid of the lapse function $(N)$ and therefore desist from admitting diffeomorphic invariance. This situation may only be circumvented under the choice $\gamma \propto N$. We therefore initiate our analysis with the choice
\be\label{gamma} \gamma = \gamma_0 N,\ee
where $\gamma_0$ is a constant. The point Lagrangian therefore reads as,
\be \label{Cov2}L_{\mathrm{cov}2} = \left[-{3{\dot z}^2\over 4{N\sqrt z}} + {9\over 4}\gamma_0 \sqrt{z}\dot z\right] f(\Phi) +\left({9\over 2}\gamma_0 z^{3\over 2}-{3\sqrt{z}\dot z\over 2N}\right)f'{\dot\Phi}+ \left({\omega(\Phi)\over N}{\dot\Phi}^2 - N U(\Phi) + {1\over 2N} {\dot\phi}^2 - N V(\phi)\right)z^{3\over 2}. \ee
One can now compare action \eqref{A2} to find that under the choice \eqref{gamma}, both $\dot N$ nor $\dot \gamma$ are absent from with the point Lagrangian \eqref{Cov2}. The corresponding canonical momenta are now,
\be\label{Mom2} p_z = -{3{\dot z}f\over 2{N\sqrt z}} + {9\over 4}\gamma_0 \sqrt{z}f - {3\sqrt{z}\over 2N}f'\dot\Phi;\;\;p_\Phi = {2\omega\over N}z^{3\over 2}\dot\Phi -{3\sqrt{z}\dot z f'\over 2N} + {9\gamma_0\over 2}z^{3\over 2}f';\;\;p_\phi = {z^{3\over 2}\over N}\dot \phi;\;\;p_N = 0.\ee
Again, since none of the momentum densities contain derivative of the lapse function, so the correlated constraint $\lambda_1 = p_N \approx 0$ strongly vanishes, and the constraint may be safely neglected. Now, from the first two relations of \eqref{Mom2} we compute,
\be \label{vel} \dot z = -{2N\sqrt{z}\over 3f}p_z + {3N\over 2}\gamma_0 z - {2N f'\over \sqrt{z} f}\left[{f p_\Phi - z f' p_z - {9\over 4}\gamma_0 z^{3\over 2}f f'\over 3f'^2 + 4 \omega f}\right];\;\dot \Phi = {2N\over z^{3\over 2}}\left[{f p_\Phi - z f' p_z - {9\over 4}\gamma_0 z^{3\over 2}f f'\over 3f'^2 + 4 \omega f}\right],\ee
and construct the Hamiltonian in a straight forward manner, which turns out to be,
\be \label{HCov}\begin{split} &\mathbb{H}_{\mathrm{Cov2}} = N\Bigg[-{\sqrt{z}\over 3f}p_z^2 + {3\gamma_0\over 2} z p_z + {p_\phi^2\over 2 z^{3\over 2}} - {27\over 16} \gamma_0^2 z^{3\over 2}f + (U + V)z^{3\over 2}\\&
+ {1\over 3f'^2 + 4 \omega f}\left({f'^2\over f} \sqrt{z} p_z^2 + {f \over z^{3\over 2}}p_\Phi^2 - {2f'\over \sqrt{z}}p_z p_\Phi + {9\over 2}\gamma_0f'^2 z p_z - {9\over 2}\gamma_0 f f' p_\Phi + {81\over 16}\gamma_0^2 z^{3\over 2} ff'^2\right)\Bigg]  = N\mathrm{H}.
\end{split}\ee
Thus, DI is established yet again for yet another non-trivial connection. The important point is to note that, apart from setting the shift vector to vanish, this connection requires to fix the unknown parameter $\gamma(t) \propto N(t)$. In this process, $\gamma(t)$ is fixed once and forever, from the fundamental physical consideration that a `Locally Lorentz invariant' scalar-vector-tensor form of GSTG theory admits `Diffeomorphic invariance' provided the parameter $\gamma \propto N$.

\subsection{The phase-space structure for action \eqref{Qcov1}~ (Connection-3):}

In view of \eqref{cterm} one may find,
\be\label{Qa3}\left(Q^\alpha - \hat{Q}^\alpha\right)f_{,\alpha}(\Phi) = -\frac{3}{N^2}\left({\dot z\over z}+{N^2\gamma \over z}\right)f'{\dot\Phi}\ee
Next, using the expression \eqref{Q3} and \eqref{Qa3}, the action \eqref{Qcov1} is expressed as,
\be\begin{split} \label{A3} A_{\mathrm{cov}3}&=\int\Bigg[ -{3{\dot z}^2\over 4{N\sqrt z}}f(\Phi)+\frac{3N{\gamma} f \sqrt z}{2}\left({\dot z\over 2z}+{\dot N\over N}\right)+\frac{3N{\dot\gamma} f \sqrt z}{2}-\frac{3}{2N}({\dot z\sqrt z}+{N^2\gamma \sqrt z})f'{\dot\Phi}\\&+\left({\omega(\Phi){\dot\Phi}^2\over N}z^{3\over 2} - Nz^{3\over 2}U(\Phi)\right)+\left({1\over 2N^2} {\dot\phi}^2 - V(\phi)\right) Nz^{3\over 2}\Bigg]dt,\end{split}\ee
where, $\phi$ is the additional scalar field required to drive inflation and $U(\Phi)$, $V(\phi)$ are the potentials associated with $\Phi$ and $\phi$ respectively. The point lagrangian therefore reads as,
\be\begin{split} \label{Cov3}L_{\mathrm{cov}3}& =\Bigg[ -{3{\dot z}^2\over 4{N\sqrt z}}f(\Phi)+\frac{3N{\gamma} f \sqrt z}{2}\left({\dot z\over 2z}+{\dot N\over N}\right)+\frac{3N{\dot\gamma} f \sqrt z}{2}-\frac{3}{2N}({\dot z\sqrt z}+{N^2\gamma \sqrt z})f'{\dot\Phi}\\&+\left({\omega(\Phi){\dot\Phi}^2\over N}z^{3\over 2} - Nz^{3\over 2}U(\Phi)\right)+\left({1\over 2N^2} {\dot\phi}^2 - V(\phi)\right) Nz^{3\over 2}\Bigg].\end{split} \ee
The corresponding canonical momenta are,
\be\begin{split}& p_z = - {3\dot z f\over 2N\sqrt z} + {3N\gamma f\over 4\sqrt z} - {3\sqrt z\over 2N}f'\dot \phi;\;\;p_\Phi =  -{3N\gamma \sqrt zf'\over 2} - {3\sqrt{z}\dot z f'\over 2N} + {2\omega z^{3\over 2}\dot \Phi\over N};\\&
p_\gamma = {3N f{\sqrt z}\over 2};\;\;\;\; p_N = {3\gamma f\sqrt z\over 2},~~~p_\phi ={\dot\phi z^{3\over 2}\over N}.\end{split}\ee
Thus, one may compute $\dot z$ and $\dot\Phi$ from the expressions of  $p_z$ and $p_{\Phi}$ respectively as,
\be\begin{split}&{\dot z} =-{2N\sqrt{z}\over 3f}p_z + {N^2\gamma\over 2}  - {z f'{\dot\Phi}\over f}\\&
= -{2N\sqrt{z}\over 3f}p_z +{N^2\gamma\over 2}  - {N f'\over \sqrt{z} f}\left[{2f p_\Phi - 2z f'p_z + {3\over 2}N\gamma \sqrt z f f'\over 3f'^2 + 4 \omega f}\right],\\&
{\dot\Phi}={N\over 2\omega z^{3\over 2}}\left[p_{\Phi}+\frac{3\dot z \sqrt z f'}{2N}+\frac{3N\gamma\sqrt zf'}{2}\right]={N\over z^{3\over 2}}\left[{2f p_\Phi - 2z f'p_z + {3\over 2}N\gamma \sqrt z f f'\over 3f'^2 + 4 \omega f}\right].\end{split}\ee
However, there exists a second class constraint $\lambda_1 = Np_N - \gamma p_\gamma \approx 0$ which is required to introduce into the constrained Hamiltonian through a Lagrange multiplier $(u_1)$, to cast the primary Hamiltonian,
\be\begin{split}\label{PH3} \mathbb{H}_{p3} =& \mathbb{H}_c + u_1(Np_N - \gamma p _\gamma) = \dot z p_z + \dot \Phi p_\Phi + \dot N p_N + \dot \gamma p_\gamma + \dot \phi p_\phi - L_{\mathrm{cov}3} + u_1(Np_N - \gamma p _\gamma)\\&=-{N\sqrt{z}\over 3f}p_z^2 + {N^2\gamma p_z\over 2}+ {Np_\phi^2\over 2 z^{3\over 2}} - {3N^3f\gamma^2\over 16\sqrt z}+ N(U + V)z^{3\over 2}\\&
+ {N\over 3f'^2 + 4 \omega f}\left({f'^2\over f} \sqrt{z} p_z^2 + {f \over z^{3\over 2}}p_\Phi^2 - {2f'\over \sqrt{z}}p_z p_\Phi\right)\\&+ {1\over 3f'^2 + 4 \omega f}\left({9\over 16\sqrt z}N^3\gamma^2  ff'^2-{3\over 2}N^2\gamma f'^2  p_z + {3N^2\gamma f f' p_\Phi\over 2z}  \right)+u_1(Np_N - \gamma p _\gamma).\end{split}\ee
Now, at this end to preserve the constraints over time, we compute the Poisson bracket, with the primary Hamiltonian to find,
\be\label{APoisson3}\begin{split}{\dot\lambda_{1}}=\left\{ \lambda_{1}, \mathbb{H}_{p3}\right\}= &-N\Bigg[-{\sqrt{z}\over 3f}p_z^2 + {p_\phi^2\over 2 z^{3\over 2}} +\frac{2N\gamma p_\gamma}{2}-\frac{9fN^2\gamma^2}{16\sqrt z}+ (U + V)z^{3\over 2}+{1\over 3f'^2 + 4 \omega f}\bigg({f'^2\over f} \sqrt{z} p_z^2\\& + {f \over z^{3\over 2}}p_\Phi^2 - {2f'\over \sqrt{z}}p_z p_\Phi-\frac{27N^2\gamma^2ff'^2}{16\sqrt z}-3N\gamma f'^2p_z+3N\gamma ff'p_{\Phi}\bigg)\Bigg]+\gamma\Bigg[{N^2\over 2}p_z\\&-{6N^3\over 16\sqrt z} \gamma f+{1\over {3f'^2 + 4 \omega f}}\bigg({9N^3\gamma ff'^2\over 8\sqrt z} ff'^2 - {3N^2\over 2}f'^2 p_z + {3N^2ff'p_{\Phi}\over 2z}\bigg)\Bigg]\approx 0,\end{split}\ee
which is again a second class constraint.\\

\noindent
\textbf{Inspection:}\\
The primary Hamiltonian is found to contain terms such as, ${\gamma zp_z}, {N^3 f \gamma^2\over \sqrt z}, {N^2\gamma f'^2 p_z \over  {3f'^2 + 4 \omega f}}, {N^2\gamma ff' p_\Phi\over {2z(3f'^2 + 4 \omega f)}}, {N^3\gamma^2 ff'^2 \over {\sqrt z(3f'^2 + 4 \omega f)}}$,  which desist from admitting diffeomorphic invariance. This situation may only be circumvented under the choice $\gamma \propto {1\over N}$. We therefore resume our analysis with the choice
\be\label{gamma3} \gamma = {\gamma_0\over N},\ee
where $\gamma_0$ is a constant. The point Lagrangian therefore reads as,
\be\begin{split} \label{MCov3}L_{\mathrm{cov}3}& =\Bigg[ -{3{\dot z}^2\over 4{N\sqrt z}}f(\Phi)+\frac{3{\gamma_0}\dot z f }{4\sqrt z}-\frac{3}{2N}{\dot z\sqrt z}f'{\dot\Phi}-{3\gamma_0 \sqrt z\over 2}f'{\dot\Phi}\\&+\left({\omega(\Phi){\dot\Phi}^2\over N}z^{3\over 2} - Nz^{3\over 2}U(\Phi)\right)+\left({1\over 2N^2} {\dot\phi}^2 - V(\phi)\right) Nz^{3\over 2}\Bigg].\end{split} \ee
The corresponding canonical momenta are,
\be\begin{split}\label{Mom3}& p_z = - {3\dot z f\over 2N\sqrt z} + {3\gamma_0 f\over 4\sqrt z} - {3\sqrt z\over 2N}f'\dot \phi;\;\;p_\Phi =  -{3\gamma_0 \sqrt zf'\over 2} - {3\sqrt{z}\dot z f'\over 2N} + {2\omega z^{3\over 2}\dot \Phi\over N};\\&
~~~p_\phi ={\dot\phi z^{3\over 2}\over N},~~~p_N=0.\end{split}\ee
Since none of the momentum densities contain derivative of the lapse function, so the correlated constraint $\lambda_1 = p_N \approx 0$ strongly vanishes, and the constraint may be safely neglected. Thus from the equations of \eqref{Mom3}, one can write
\be\begin{split}&{\dot z} =-{2N\sqrt{z}\over 3f}p_z + {N\gamma_0\over 2}  - {z f'{\dot\Phi}\over f}
= -{2N\sqrt{z}\over 3f}p_z +{N\gamma_0\over 2}  - {N f'\over \sqrt{z} f}\left[{2f p_\Phi - 2z f'p_z + {3\over 2}\gamma_0 \sqrt z f f'\over 3f'^2 + 4 \omega f}\right],\\&
{\dot\Phi}={N\over 2\omega z^{3\over 2}}\left[p_{\Phi}+\frac{3\dot z \sqrt z f'}{2N}+\frac{3\gamma_0\sqrt z f'}{2}\right]={N\over z^{3\over 2}}\left[{2f p_\Phi - 2z f'p_z + {3\over 2}\gamma_0 \sqrt z f f'\over 3f'^2 + 4 \omega f}\right],\end{split}\ee
and construct the Hamiltonian in a straight forward manner, which is given by,
\be\begin{split}\label{MPH3} \mathbb{H}_{\mathrm{Cov3}} =& \dot z p_z + \dot \Phi p_\Phi + \dot \phi p_\phi - L_{\mathrm{cov}3} \\&=N\bigg[-{\sqrt{z}\over 3f}p_z^2 + {\gamma_0 p_z\over 2}+ {p_\phi^2\over 2 z^{3\over 2}} - {3f\gamma_0^2\over 16\sqrt z}+ (U + V)z^{3\over 2}\\&
+ {1\over 3f'^2 + 4 \omega f}\left({f'^2\over f} \sqrt{z} p_z^2 + {f \over z^{3\over 2}}p_\Phi^2 - {2f'\over \sqrt{z}}p_z p_\Phi+ {9\over 16\sqrt z}\gamma_0^2  ff'^2-{3\over 2}\gamma_0 f'^2  p_z + {3\gamma_0 f f' p_\Phi\over 2z}  \right)\bigg] = N\mathrm{H}.\end{split}\ee
DI is thus established for connection-3 too and since the Hamiltonian contains momentum upto quadratic terms, so it is well-behaved, and would lead to modified Wheeler-deWitt equation upon quantization. Semiclassical formulation is also possible and Hartle criteria supposedly would be admissible, which we pose to consider in future.

\section{Conclusion:}

Extensive interest in `teleparallel gravity theories' is currently in the wane due to the exploration of their pathological behaviour. Initially `generalized symmetric telerarallel gravity' (GSTG) was thought to be free from the issues encountered in `generalized metric telerarallel gravity' (GMTG) \cite{LLT,fT1l,fT2}. Nonetheless, GSTG theory also primarily was found to be diseased with the violation of `Local Lorentz Invariance' (LLI). Fortunately, LLI has been established in the covariant formulation of equivalent scalar-tensor formulation of GMTG, taking spin connection into account \cite{Tcov1, Tcov2} and covariant GSTG taking affine connection into account \cite{Hoh1, Hoh2, Qcov2}. However, covariant GSTG has also been formulated with scalar-nonmetricity coupling in scalar-vector-tensor form \cite{Qcov1, Hu}. Unfortunately, both the theories are plagued with coupling issues and ghost degree of freedom. To be specific, in the scalar-non-metricity formulation of GSTG, the scalar mode appears with a negative kinetic energy and hence is a ghost. This extra scalar degree of freedom does not admit linear perturbation around maximally symmetric cosmological backgrounds, indicating strong coupling issue \cite{fT2,fQ1,fQ2,fQ3}. Nonetheless, this fact do not seal the fate of the `teleparallel theories of gravity'. Firstly, there is no reason to rely simply on linear perturbation. One has to compute higher-order perturbation before making a conclusive remark, which is of-course extremely difficult. Next, it is shown that the problem with Ostrogradski ghost may be alleviated by associating a non-minimally coupled scalar field, even at the linear perturbation level \cite{Hu2}. There is yet another issue which is no less important. In the presence of ghost degrees of freedom, Diffeomorphic invariance (DI) breaks down \cite{H2,H3,H4}. However, all these works are based on non-covariant scalar-tensor equivalent action and as such the authors \cite{H2,H3,H4} sacrificed local Lorentz invariance, and ends up with the loss of diffeomorphism invariance. Considering the LLI action, if the theory turns out to be free of ghosts, then all the pathologies would be alleviated. In these respect, it is required to study GSTG theory earnestly. In the present manuscript, we have attempted to formulate the phase-space structure of GSTG theory with non-coincident gauge, of-course for a specific minisuperspace of cosmological interest, being devoid of the shift vector. In the following we brief our findings.\\

1. The LLI GSTG theory may be constructed if in addition to the metric, affine connections are also treated as independent variables. For the spatially flat $(k = 0)$ isotropic and homogeneous RW metric, three sets of non-vanishing connections are realizable. One very important outcome of the present work is that, except for the first, the other two sets of connection comprising of an unknown time-dependent parameter $\gamma$, admit linear non-metricity theory of gravity - $f_{QQ} = 0$, while the connection variation equations identically vanish. This means, these two connections either give nothing other than GTR, or may be applied in the linear scalar-vector-tensor form of GSTG theory, in which the non-metricity scalar is linearly coupled to arbitrary functional form of a scalar field. \\

2. We therefore construct the phase-space structure for the `auxiliary scalar-tensor' GSTG action $(f_{QQ} \ne 0)$ with the first set of connections (connection-1), for which the connection variation equation is trivially satisfied. The Dirac-Bergamann analysis of such a theory earlier revealed that the auxiliary scalar together with the shift vector act as non-propagating variables carrying $1\over 2$ degree of freedom and are responsible for the violation of diffeomorphic invariance (DI) in general \cite{H2,H3,H4}. As a result, Hamiltonian could not be constructed. Nonetheless, it has been noticed that the appearance of ghost and the breakdown of DI essentially are artefact of the shift vector, since the momentum associated with the auxiliary scalar field involves the first derivative of the shift vector. Clearly, in a minisuperspace model of cosmological interest, such as the Robertson-Walker metric which is devoid of the shift vector, the theory must not be plagued with such pathologies. Applying Dirac-Bergamann algorithm, we have explicitly exhibited that indeed DI remains preserved. Lamentably, the structure of the Hamiltonian turns out to be frightening, as it contains momenta in fourth degree. Further, the minimally coupled scalar field we considered to drive inflation, is now coupled with the derivative of the function $f(Q)$, which is vexatious. In a nut-shell, the eerie Hamiltonian cannot be maneuvered and the study of inflation in view of classical field equations is not justified. Thus, GSTG theory for $f_{QQ} \ne 0$ has actually no merit.\\

3. Next we consider the covariant action for the linear `scalar-vector-tensor' formulation of GSTG theory taking into account all the three sets of connections (connection-1, connection-2, connection-3) for which the affine connections are non-vanishing in the flat Robertson-Walker metric under consideration. However, connection variation equations for all the three cases trivially vanish. Connection-1 is the simplest one for which the non-metricity scalar is identical to the one that appears in the coincidence gauge $(\Gamma = 0)$. In this case, the system is devoid of constraints, since the only one, viz., $\lambda_1 = p_N \approx $ strongly vanishes and the Hamiltonian is constructed in a straight forward manner. DI remains preserved resulting in a well-behaved Hamiltonian suitable for quantization and semiclassical approximation, that could easily meet `Hartle criteria'. Connection-2 contains an arbitrary functional parameter $\gamma(t)$. Following Dirac-Bergamann constraint analysis, it is revealed that even in the absence of the shift vector DI can only be established imposing an additional condition such as $\gamma(t) \propto N(t)$, where, $N$ is the lapse function. In fact, earlier some authors had chosen $\gamma(t)$ arbitrarily to explore cosmological solutions \cite{Avik, Dima}. In this respect this is a unique result, since from the physical consideration $\gamma(t)$ is fixed once and forever. Once the choice $\gamma(t) \propto N(t)$ is imposed, the system becomes devoid of constraints as before, and the Hamiltonian is constructed in a straight forward manner. Finally we take up connection-3, follow Dirac-Bergamann constraint analysis and find that $\gamma(t) \propto N^{-1}(t)$ is the condition required to establish DI. Once again, upon imposing the condition, Hamiltonian is constructed in a straight forward manner as before without requiring constraint analysis. For both the connections 2 and 3, the Hamiltonian is quadratic in momenta and therefore well-behaved. Quantization is straight forward and semiclassical approximation is viable to meet `Hartle criteria'. Consequently, there happens to be no extra degree of freedom and all the pathologies including the ghost disappear automatically. \\

In a nut-shell, it is revealed that description of gravity in view of non-metricity practically has no merit for $f(Q)$ theory, with $(f_{QQ} \ne 0)$. However, $f(R,Q)$ theory would be free from all these problems as it has been exhibited for $f(R,\mathrm{T})$ theory \cite{MKA}. On the contrary, the scalar-vector-tensor GSTG theory in which the non-metricity scalar is linearly coupled to arbitrary function of a scalar field is well-behaved, being free from uncanny (ghost) degrees of freedom and therefore requires further investigation. These include, canonical quantization, semiclassical approximation, the study of inflation and to explore the the late-stage of cosmic evolution in the radiation as well as in the matter (pressure-less dust) dominated eras. Additionally, it should be mentioned that there exists yet another set of connections (connection-4) for non-flat $(k \ne 0)$ case, which also requires investigation. These problems we pose in the future.

\end{document}